\documentclass[%
 reprint,
longbibliography,
superscriptaddress,
 amsmath,amssymb,
]{revtex4-2}

\usepackage{xcolor}
\def\AJ#1{{\color{black}{#1}\color{black}}}
\def\LH#1{{\color{black}{#1}\color{black}}}

\usepackage{soul} 
\usepackage{graphicx}
\usepackage{dcolumn}
\usepackage{bm}
\usepackage{hyperref}

\usepackage[utf8]{inputenc}
\usepackage[T1]{fontenc}
\usepackage{lmodern}

\usepackage{makecell}
\usepackage{amsmath,multirow}
\usepackage{mathtools}
\usepackage{relsize}
\usepackage{array}

\usepackage{mathtools} 
\usepackage[caption=false,labelformat=simple]{subfig}

\begin{document}

\title{Symmetric conformity functions make decision-making processes independent of the distribution of learning strategies}

\author{Arkadiusz J\k{e}drzejewski}
\thanks{EUTOPIA-SIF COFUND Marie Skłodowska-Curie Fellow at CY Cergy Paris Université, CY Advanced Studies, Neuville-sur-Oise, 95000, France}
\email{arkadiusz.jedrzejewski@cyu.fr}
\affiliation{Laboratoire de Physique Th\'{e}orique et Mod\'{e}lisation, CY Cergy Paris Universit\'{e}, CNRS, F-95302 Cergy-Pontoise, France}

\author{Laura Hern\'andez}%
\email{laura.hernandez@cyu.fr}
\affiliation{Laboratoire de Physique Th\'{e}orique et Mod\'{e}lisation, CY Cergy Paris Universit\'{e}, CNRS, F-95302 Cergy-Pontoise, France}


\date{\today}

\begin{abstract}
Two main procedures characterize the way in which social actors  evaluate the qualities of the options in decision-making processes:  they either seek to evaluate their intrinsic qualities (individual learners), or they rely on the opinion of the others (social learners). For the latter, social experiments have suggested that the mathematical form of the probability of adopting  an option, called the \textit{conformity function}, is symmetric in the adoption rate. However, the literature on decision-making includes models where social learners employ either symmetric or non-symmetric conformity functions. \LH{We generalize a particular case studied in a previous work, and we show analytically that if the conformity function is symmetric,  the details of  the probability distribution  of the propensity of the agents to behave  as a social or an individual learner do not matter, only  its expected value influences the determination of the steady state. We also show that in this case, the same steady state is reached for two extreme dynamical processes:  one that considers  \AJ{propensities} as  idiosyncratic properties of the agents (each agent being individual learner always with the same probability) and the opposite one, that allows them to change their \AJ{propensity}  during the dynamics. This is not the case   if the conformity function is non-symmetric.} This fact can inspire experiments that could shed light on the debate about \LH{mathematical properties of conformity functions}.\\\\
Post-print of \href{https://doi.org/10.1103/PhysRevResearch.6.033093}{Phys. Rev. Research \textbf{6}, 033093 (2024)}.\\
Copyright (2024) by the American Physical Society.
\end{abstract}

\maketitle

\section{\label{sec:intorduction}Introduction}

Decision making is an individual task that benefits from detailed knowledge about the possible options. 
The vast literature addressing the way in which different species of animals, and in particular humans, acquire this knowledge is pluri-disciplinary and targets different aspects of the problem \cite{Ren:etal:11,Apl:etal:14,Ken:etal:18}. 
Social actors are usually classified according to their learning strategies as \textit{individual learners}, those who search to identify the intrinsic merits of the options without suffering any peer pressure, or \textit{social learners}, those who simply follow their peers' choice. However, this is a rough classification as each class entails a  variety of cognitive processes that are very difficult to disentangle experimentally. Early studies on decision-making were challenged by new experimental techniques~\cite{Whi:Ham:92,Whi:etal:09}. A  question that raised strong debates concerns the transmission of learning abilities in the light of natural selection. As social learning (in any of its forms) is considered less costly than individual learning, it is supposed to enhance individual fitness and then prevail~\cite{Boy:Ric:88}. However, A. Rogers showed that this may not be the case if the environment is subject to changes. In this case, if social learners are selected, their proportion in society increases, and the probability that they obtain wrong information about the environment by copying other social learners with ``old'' information increases, and therefore, their fitness diminishes. This is known as Rogers' paradox~\cite{Rog:88}. 
Rogers' paradox does not mean that social learning---thus culture---prevents social agents from adapting to the environment, it just points out that a model that only evaluates the cost-benefit of the chosen strategies is not enough to account for the observations. Rogers himself had suggested to introduce some biases in the social learning, like copying preferentially high fitness individuals, or as proposed by Boyd and Richerson,  copying only individual learners.  None of them lifted the paradox, as the fitness of the group decreases with generations because the strategy of social learners is frequency dependent while that of individual learners is not~\cite{Boy:Ric:95}. Other modifications introduce the possibility for the strategies of an individual to evolve according to different situations (cost of individual learning, changing environment, fitness of the neighbours, etc.). These modifications may lift Rogers' paradox or not depending on details of the parameters~\cite{Enq:Eri:Ghi:07,Ehn:Lal:12}. All this shows that the problem of how learning strategies are transmitted goes beyond a cost-benefit problem and that flexibility in the learning strategies is essential in order to maintain a high fitness of the population. 

Another aspect of the problem would be to ask how a given generation  composed of individual and social learners reaches a collective decision.  In this case, the particular ways in which both social and individual learners acquire new knowledge are studied. For example, social learners conform to an option because of peer pressure. 
\LH{On the other hand, individual learners  seek information about the options on their own. This does not necessarily imply that they ignore the choice of others. They may also take advantage of it by including this information in their evaluation criterion, as has been studied in \AJ{Refs.}~\cite{Dim:Gar:12, Bal:Pei:18}. The effect of the global preferences of the population can be considered either positively or negatively by individual learners, depending on the context.}

\LH{A positive global effect occurs when a product or service becomes more valuable as more people use it.}
Social media platforms or e-commerce marketplaces manifest positive  \LH{global} effects. 
More users create more content, while more sellers provide a wider selection of products.
Consequently, the value of the platform increases.
Positive \LH{global} effects can also reinforce pro-environmental behaviors for which it is critical to lower the behavioral difficulty of engagement in order to increase adoption rates \cite{Byr:etal:16}.
For instance, as the popularity of electric vehicles increases, more investments in charging infrastructure are expected, facilitating the adoption for new users. \LH{It should be noticed that here, \AJ{the mechanism behind the positive global effect} is different from mere peer pressure, where the individual feels the need to behave as their \AJ{peers}.}
On the other hand, a negative \LH{global}  effect arises if the value of a product or service decreases as more people use it.
This can occur when high demand causes congestion, reduce availability, or lower the quality of the service \cite{Bal:Pei:18}. 
For example, public transportation may face overcrowding, while a bike-sharing system may suffer from bike shortages. 

\LH{With these considerations we see that the merit of an option is not necessarily constant, as in Ref. \cite{Yan:etal:21}, but may depend on the adoption rate. }
It is interesting to notice here that this point also addresses the main ingredient of  Rogers' paradox~\cite{Boy:Ric:95}.

Recently, a  dynamical system model has showed that social learners may, when numerous, impair collective performance~\cite{Yan:etal:21}. This model considers the proportion of social and individual learners as well as the value of the merit as a fixed parameter and chooses a symmetric \textit{conformity function} to describe  social learning.  However, the specific form of social learning function is still a subject of study. McElreath et al. have detailed different heuristics for social learning which lead to either symmetric or non-symmetric functional forms~\cite{Mce:etal:08}.   
Experiments, both in laboratory and in real settings, \LH{reveal} different ways in which  individuals learn from peers~\cite{Cla:etal:14} and whether there is a general mathematical form representing a general social learning function is far from being clear~\cite{Der:God:Ray:13,Mor:etal:12}. These experiments also observed the situation where subjects change their strategies during the experiment~\cite{Mor:etal:12} alternating the ways in which they gather information (either by learning individually or by getting information from their peers). 

\LH{In this work, we present a general theoretical  model that allows us to explore analytically the possible  outcomes of  dynamics
where the social learning function may be symmetric, as the one considered  in Ref.~\cite{Yan:etal:21}, or non-symmetric, as in the well-known $q$-voter model~\cite{Cas:Mun:Pas:09}. The model also allows for a general distribution  of  such strategies, \AJ{$\phi(p_i)$}, where $p_i$ is the probability that a given agent acts  as an individual learner, \AJ{otherwise it acts as a social learner}. Moreover, we explore the two possible extreme cases for the evolution of strategies with time, either each agent has a fixed probability of being an individual learner, initially chosen from the distribution  $\phi(p_i)$ \textit{(quenched dynamics)}, or its strategy can evolve in the same time scale as the choice of the options. 
\AJ{In this case, the agent chooses its probability $p_i$ of acting as an individual learner from the distribution $\phi(p_i)$ at each time step \textit{(annealed dynamics)}. }
}

\LH{Our main general result points at the consequences of the conformity function being  symmetric.   We prove analytically that if the social learners use a symmetric conformity function, then the steady state of the system does not depend on  $\phi(p_i)$, but only on its first moment, $\bar{p}$, and this is true regardless of the form of the individual learners' function. We illustrate this general result by considering particular forms of social and individual learning functions as well as a particular distribution of these learning strategies. We  show that  the phase diagrams of the quenched and annealed dynamics are identical for the symmetric conformity function case, while they differ when the conformity function is non-symmetric. The differences may involve  the presence of discontinuous or continuous transitions depending on the  parameters.}

Our results confirm and extend those presented in Ref.~\cite{Yan:etal:21}, which is a particular case of our model.

\section{Model}
\begin{figure*}[!t]
	\centering
	\includegraphics[width=\linewidth]{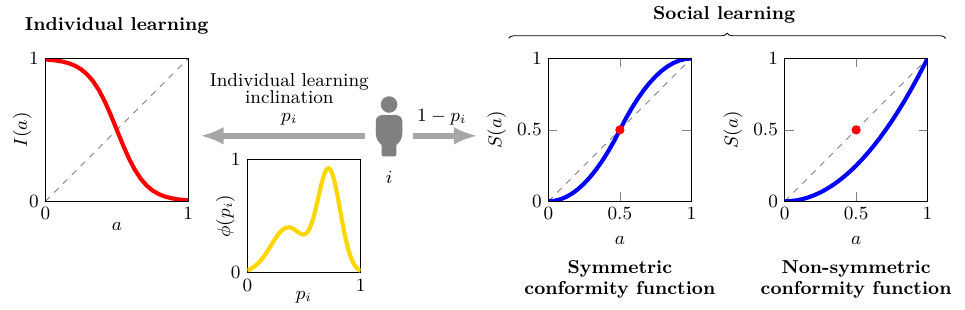}
	\caption{\label{fig:model-diagram} 
	Model diagram: agent $i$ has a personal inclination towards individual learning, represented by the probability $p_i$ of choosing this strategy. 
    \AJ{With complementary probability, $1-p_i$, social learning is employed. 
    Based on the chosen strategy, a decision is made using the corresponding learning function, either $I(a)$ or $S(a)$, according to the rules outlined in Table~\ref{tab:prob}.
    Both learning functions may depend on the current adoption rate of the option, $a$.
    We categorize conformity functions into two groups: symmetric, where the graph remains unchanged after a $180^\circ$ rotation around the point $S(0.5)=0.5$, and non-symmetric.} 
    In the quenched dynamics, $p_i$ is \LH{initially} assigned to agent $i$ \LH{ and remains fixed during the evolution}, whereas in the annealed dynamics, $p_i$ is assigned to agent $i$ \LH{at each time step. In both cases, $p_i$ is drawn from distribution $\phi(p_i)$}. 
	}
\end{figure*}

We study a situation where individuals (agents) of the society  have to make a binary decision \AJ{between adopting ($A$) or rejecting ($B$)  \LH{a certain} option.
This option may represent a product, a behaviour, or a social norm.
The agents, labeled by index $i$, are randomly selected in sequence to make their decisions.
An agent behaves as an individual learner with probability $p_i$. 
Otherwise, it behaves as a social learner, which happens with complementary probability $1-p_i$. 
The values of $p_i$ come from a general distribution, \AJ{$\phi(p_i)$}, as illustrated in the left part of Fig.~\ref{fig:model-diagram}. 
There are two possible dynamics: either the values of $p_i$ are assigned to the agents from the start and stay with them unchanged throughout the entire process (quenched dynamics), or the values of $p_i$ are reassigned to the agents each time they are selected to make a decision  (annealed dynamics).
}

An agent that follows the individual \AJ{learning strategy evaluates the probability of adopting the option using function $I(a)$, where $a$ represents the fraction of the population that has adopted that option. 
The probability of rejecting the option is given by the complementary probability $1-\AJ{I(a)}$.}
\AJ{
This function reflects how the perceived utility of an option changes with the evolving fraction of adopters in the population.
Such changes are caused by the advantages or disadvantages directly resulting from the size of the group of adopters that impact agents' payoffs or risk-adjusted returns \cite{Dim:Gar:12,Bal:Pei:18}. 
However, it is important to note that this function does not account for social influence arising from conformity and the willingness to adhere to social norms---factors that do not affect the intrinsic value of the option.}
\AJ{
In a particular case, the individual learning function may \LH{be a constant}, i.e., $I(a)=m$.
In this context, parameter $m$ represents the merit of adopting the option that remains unaffected by the number of adopters, as considered in Ref.~\cite{Yan:etal:21}.
}

On the other hand, \AJ{social learning arises from the observation of peers and resulting social influence.
It can take different forms described mathematically by models of cultural transmission  \cite{Ren:etal:11,Ken:etal:18} or the social impact theory \cite{Lat:81}.} 
We represent it as a conformity function, $S(x)$, that gives the probability of changing an agent's choice to the choice shared by a fraction $x$ of the agent's peers.
Based on mathematical models \cite{Boy:Ric:88,Lat:81} and experimental studies \cite{Apl:etal:15,Cla:etal:14,Cla:Bow:Whi:12,Cla:Whi:12,Eff:etal:08,Mce:etal:08,Mor:etal:12,Mor:Lal:12} that try to characterise conformity functions, we distinguish between two classes.
Functions falling within the first class are symmetric around their midpoint $S(0.5)=0.5$, see the right part of Fig.\ref{fig:model-diagram}.
Such functions satisfy the following equation 
\begin{equation}
    \label{eq:symmetry}
    S(x)+S(1-x)=1
\end{equation}
for all $x\in[0,1]$.
All other functions are classified as non-symmetric ones.

Table~\ref{tab:prob} summarises the way individual and social learning functions are used to update the agents' \AJ{choices}.
\begin{table}[t]
\caption{\label{tab:prob} Probabilities that an agent with a given option \LH{keeps}  or changes it to the alternative one using a corresponding learning strategy.
}
\centering
\renewcommand{\arraystretch}{1.5}
\begin{tabular}{ccccc}
\toprule
\multirow{3}{*}{\parbox[t]{1.4cm}{Option before learning}}                 & \multicolumn{4}{c}{\parbox[t]{6cm}{ Probability of the option after learning}}              \\
\cline{2-5}
               & \multicolumn{2}{c}{Individual learning}          & \multicolumn{2}{c}{Social learning}                  \\ \cline{2-5}
 & \parbox[t]{1.5cm}{$A$}                        & \multicolumn{1}{c}{\parbox[t]{1.5cm}{$B$}} & \parbox[t]{2cm}{$A$}                            & \multicolumn{1}{c}{\parbox[t]{1.5cm}{$B$}} \\\colrule
$A$                                   & \multicolumn{1}{c}{\AJ{$I(a)$}} & \multicolumn{1}{c}{$1-\AJ{I(a)}$}                & \multicolumn{1}{c}{$1-\AJ{S(1-a)}$} & \multicolumn{1}{c}{\AJ{$S(1-a)$}}                \\
$B$                                   & \multicolumn{1}{c}{\AJ{$I(a)$}} & \multicolumn{1}{c}{$1-\AJ{I(a)}$}                & \multicolumn{1}{c}{\AJ{$S(a)$}}     & \multicolumn{1}{c}{$1-\AJ{S(a)}$}               \\ \botrule
\end{tabular}
\end{table}

\section{Results}
We study two dynamical scenarios where $p_i$ is either a quenched or annealed property of  agent $i$~\cite{Jed:Szn:19, Jed:Szn:20}. 
\AJ{In section~\ref{sec:gen},} we present the general \AJ{results} for these two dynamics without imposing any specific form of the \AJ{learning functions, $I(a)$ and $S(a)$, and the} learning strategy distribution, \AJ{$\phi(p_i)$}.  
\AJ{In section~\ref{sec:case},} we illustrate the behaviour of the system \AJ{with specific forms of these functions, which are  discussed in \LH{the} literature \cite{Yan:etal:21,Jed:Szn:19}.
Our general results are summarized in Table~\ref{tab:sum-v1}.}

\subsection{General results} 
\label{sec:gen}
Let \AJ{$\phi(x)$} be an arbitrary distribution with mean $\bar{p}=\int x\AJ{\phi(x)}dx$. 
\AJ{Additionally, let $I(a)$ and $S(a)$ be arbitrary functions returning probabilities that influence the choices of agents in the way presented in Table~\ref{tab:prob}.}
\subsubsection{Annealed dynamics}
\label{sec:gen-ann}

At each time step, each agent is assigned a probability of being individual learner, \AJ{$p_i$}, from the distribution \AJ{$\phi(p_i)$}. It should be noticed that the probability distribution itself does not change in time.

The time evolution of the fraction \LH{$a$} of adopters of choice $A$ is given by:
\begin{equation}
\label{eq:rate-equation}
    \frac{da}{dt}=P_{B\rightarrow A}(1-a)-P_{A\rightarrow B}a,
\end{equation}
where $P_{B\rightarrow A}$ and $P_{A\rightarrow B}$ are the transition probabilities. The transition probabilities from one option to the other are formed at each step by those agents that learnt about the options either individually or socially:
\begin{equation}
\label{eq:transition-rates}
\begin{split}
    P_{B\rightarrow A}=&\int x\AJ{I(a)\phi(x)}dx+\int(1-x)\AJ{S(a)\phi(x)}dx,\\
    P_{A\rightarrow B}=&\int x\left[1-\AJ{I(a)}\right]\AJ{\phi(x)}dx\\&+\int(1-x)\AJ{S(1-a)\phi(x)}dx.
\end{split}  
\end{equation}
Having integrated the above formulas, we get:
\begin{equation}
\label{eq:transition-rates-2}
\begin{split}
    P_{B\rightarrow A}&=\bar{p}\AJ{I(a)}+(1-\bar{p})\AJ{S(a)},\\
    P_{A\rightarrow B}&= \bar{p}\left[1-\AJ{I(a)}\right]+(1-\bar{p})\AJ{S(1-a)}.
\end{split}  
\end{equation}
From Eqs.~(\ref{eq:rate-equation}) and (\ref{eq:transition-rates-2}), we get that
\begin{equation}
    \frac{da}{dt}=\bar{p}\left[\AJ{I(a)}-a\right]+(1-\bar{p})\left[(1-a)\AJ{S(a)}-a\AJ{S(1-a)}\right].
\label{eq:rate-equation_bis}
\end{equation}
Let $a^*$ denote the fixed points, which make
\begin{equation}
    \left.\frac{da}{dt}\right\vert_{a^*}=0.
\end{equation}
The fixed points satisfy the following equation: 
\begin{equation}
\label{eq:fixed}
    \bar{p}=\frac{a^*\left[\AJ{S(a^*)+S(1-a^*)}\right]-\AJ{S(a^*)}}{a^*\left[\AJ{S(a^*)+S(1-a^*)}\right]-\AJ{S(a^*)}+\AJ{I(a^*)}-a^*}.
\end{equation}
Additionally, if $I(1/2)=1/2$, we get $a^*=1/2$  for any value of $\bar{p}$. 
If the conformity function, \AJ{$S(a)$}, is symmetric, we can use Eq.~(\ref{eq:symmetry}) to simplify the above formula.
As a result, Eq.~(\ref{eq:fixed}) becomes:
\begin{equation}
    \label{eq:p-ann}
    \bar{p}=\frac{a^*-\AJ{S(a^*)}}{\AJ{I(a^*)-S(a^*)}}.
\end{equation}

\begin{table*}
\caption{\label{tab:sum-v1} Summary of the results. Formulas for the fixed points with majoritarian option of the models with different dynamics and types of conformity functions.
}
\centering
\renewcommand{\arraystretch}{1.5}
\begin{tabular}{ccc}
\toprule
    Conformity function & Annealed dynamics & Quenched dynamics\\
\colrule
\multicolumn{1}{c}{\parbox[b]{2.8cm}{Symmetric}}     &    \multicolumn{2}{c}{\parbox[t]{14.6cm}{\raggedright
Annealed and quenched dynamics lead to the same fixed points that do not depend on the full distribution of learning strategies, but only on its mean: 
\begin{equation}
    \bar{p}=\frac{a^*-\AJ{S(a^*)}}{\AJ{I(a^*)-S(a^*)}}, \text{ where } \bar{p}=\int x\AJ{\phi(x)}dx.\nonumber
\end{equation}
}} \\  \colrule
\multicolumn{1}{c}{
\parbox[t]{2.8cm}{Non-symmetric}
} &   \multicolumn{1}{c}{\parbox[t]{7.8cm}{\raggedright
Fixed points depend on the mean of the distribution of learning strategies:
\begin{equation}
    \bar{p}=\frac{a^*\left[\AJ{S(a^*)+S(1-a^*)}\right]-\AJ{S(a^*)}}{a^*\left[\AJ{S(a^*)+S(1-a^*)}-1\right]+\AJ{I(a^*)}-\AJ{S(a^*)}},\nonumber
\end{equation}
where 
\(\displaystyle\bar{p}=\int x\AJ{\phi(x)}dx.\)\newline
}}
&  
\multicolumn{1}{c}{\parbox[t]{6.8cm}{\raggedright
Fixed points depend on the whole shape of the distribution of learning strategies:
\begin{center}
$
\begin{aligned}
    a^*&=\int a_x^*\AJ{\phi(x)}dx,\text{ where}\\
    a_x^*&=\frac{x\AJ{I(a^*)}+(1-x)\AJ{S(a^*)}}{x+(1-x)\left[\AJ{S(a^*)+S(1-a^*)}\right]}.
\end{aligned}
$
\end{center}
}}
\\\botrule
\end{tabular}
\end{table*}

\subsubsection{Quenched dynamics}
\AJ{Each agent is assigned a probability of being individual learner, $p_i$,  from the distribution $\phi(p_i)$ at the beginning of the dynamics, and it keeps the same probability during the whole time evolution.}

\AJ{We divide the entire population into groups of agents who share the same value of $p_i$.
The fraction of agents with $p_i=x$ is given by:
\begin{equation}
\label{eq:division}
    \lim_{\Delta x\to 0} \int_{x}^{x+\Delta x}\phi(y)dy=\lim_{\Delta x\to 0} \phi(x)\Delta x=\phi(x)dx.
\end{equation}}

\AJ{
Let $a_x$ denote the fraction of agents who \LH{adopt} the option \LH{$A$} among those with $p_i=x$.
Consequently, $1 - a_x$ is the fraction of agents who \LH{reject} the option within the same group.
With our division given by Eq.~(\ref{eq:division}), the fraction of individuals who favor $A$ in the entire system can be expressed as a continuous analogue of a weighted average:
\begin{equation}
\label{eq:apopulation_que}
    a=\int a_x\phi(x)dx,
\end{equation}
where $a_x$ is averaged with the weights given by $\phi(x)$.}

For each value of the probability $x$, we have the following equation for the \AJ{adoption rate}:
\begin{equation}
    \label{eq:rate-que}
    \frac{da_x}{dt}=P^x_{B\rightarrow A}(1-a_x)-P^x_{A\rightarrow B}a_x,
\end{equation}
where $P^x_{B\rightarrow A}$ and $P^x_{A\rightarrow B}$ are the transition probabilities for the group of agents with  $p_i=x$:
\begin{equation}
\begin{split}
\label{eq:transition-rates-que}
    P^x_{B\rightarrow A}&=x\AJ{I(a)}+(1-x)\AJ{S(a)},\\
    P^x_{A\rightarrow B}&= x\left[1-\AJ{I(a)}\right]+(1-x)\AJ{S(1-a)}.
\end{split}  
\end{equation}

We look for the fractions of adopters among agents with $p_i=x$, $a_x^*$, that make the evolution of all the populations stationary:
\begin{equation}
    \label{eq:dax}
    \left.\frac{da_x}{dt}\right\vert_{\{a_x^*\}}=0.
\end{equation}
The fixed points are determined by combining Eqs.~(\ref{eq:rate-que}), (\ref{eq:transition-rates-que}), and (\ref{eq:dax}):
\begin{equation}
\label{eq:quenched-frac}
    a_x^*=\frac{x\AJ{I(a^*)}+(1-x)\AJ{S(a^*)}}{x+(1-x)\left[\AJ{S(a^*)+S(1-a^*)}\right]},
\end{equation}
where
\begin{equation}
\label{eq:apopulation}
    a^*=\int a_x^*\AJ{\phi(x)}dx.
\end{equation}
Additionally, if $I(1/2)=1/2$, $a_x^*=1/2$ satisfies Eq.~(\ref{eq:dax}) for any distribution $\phi(x)$. 

The fixed values of $a$ are obtained by solving Eq.~(\ref{eq:apopulation}) with the obtained formulas for $a_x^*$.
If the conformity function is non-symmetric, we cannot perform further calculations without knowing the exact distribution \AJ{$\phi(x)$}.
However, if the conformity function is symmetric, we can perform the integration in Eq.~(\ref{eq:apopulation}) without imposing any special form of \AJ{$\phi(x)$} since we can use Eq.~(\ref{eq:symmetry}) to simplify Eq.~(\ref{eq:quenched-frac}).
In such a case, we get
\begin{equation}
    \label{eq:a-que}
    a^*=\bar{p}\AJ{I(a^*)}+(1-\bar{p})\AJ{S(a^*)},
\end{equation}
so the fixed points depend only on the mean of the distribution $\bar{p}=\int x\AJ{\phi(x)}dx$.
By transforming Eq.~(\ref{eq:a-que}), we get
\begin{equation}
    \bar{p}=\frac{a^*-\AJ{S(a^*)}}{\AJ{I(a^*)-S(a^*)}}.
\label{gen_ptfix_symm_quenched}
\end{equation}

\subsubsection{Summary of general results}
\AJ{
Based on the above calculations, we can draw the following general conclusions, which are also summarised in Table~\ref{tab:sum-v1}:
\begin{itemize}
    \item If the   conformity function is symmetric, the fixed points of the quenched and the annealed dynamics are the same, and they only depend on the mean of the learning strategy distribution, $\bar{p}=\int x\phi(x)dx$, not on the details of this distribution; compare Eq.~(\ref{eq:p-ann}) with Eq.~(\ref{gen_ptfix_symm_quenched}).
    \item If the conformity function is non-symmetric, the quenched and annealed dynamics lead to different fixed-point diagrams. For the annealed dynamics, the fixed points depend only on the mean of the learning strategy distribution; see Eq.~(\ref{eq:fixed}). However, for the quenched dynamics, the actual shape of this distribution matters; see Eqs.~(\ref{eq:quenched-frac}) and (\ref{eq:apopulation}).
\end{itemize}
}
\begin{figure}[!t]
    \subfloat{\label{fig:fixed-points-k:a}}
	\subfloat{\label{fig:fixed-points-k:b}}
	\centering
	\includegraphics[width=\linewidth]{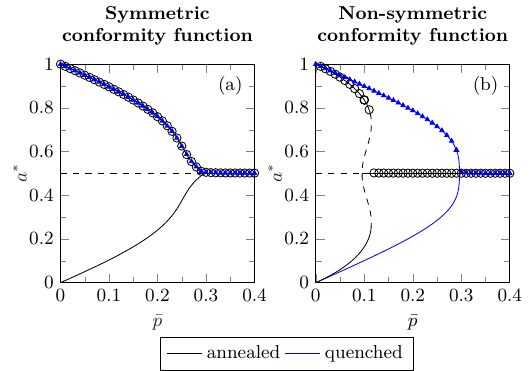}
	\caption{\label{fig:fixed-points-k} \AJ{Illustration of how different types of dynamics impact the fixed-point diagram for the model with a symmetric and non-symmetric conformity function. We present the case where $q=3$ and $k=-15$.
    Stable fixed points are represented by solid lines, while unstable ones are shown with dashed lines. }
    For the symmetric conformity function, annealed and quenched approaches produce the same diagrams.
    Symbols represent the results from the simulations of the model with $N=10^5$ agents under $\circ$ annealed and $\blacktriangle$ quenched dynamics.
    Detailed information about the simulations can be found in Appendix~\ref{sec:simulations}.
	}
\end{figure}
\begin{figure*}[!t]
\subfloat{\label{fig:fixed-points-dist:a}}
	\subfloat{\label{fig:fixed-points-dist:b}}
 \subfloat{\label{fig:fixed-points-dist:c}}
	\centering
	\includegraphics[width=\linewidth]{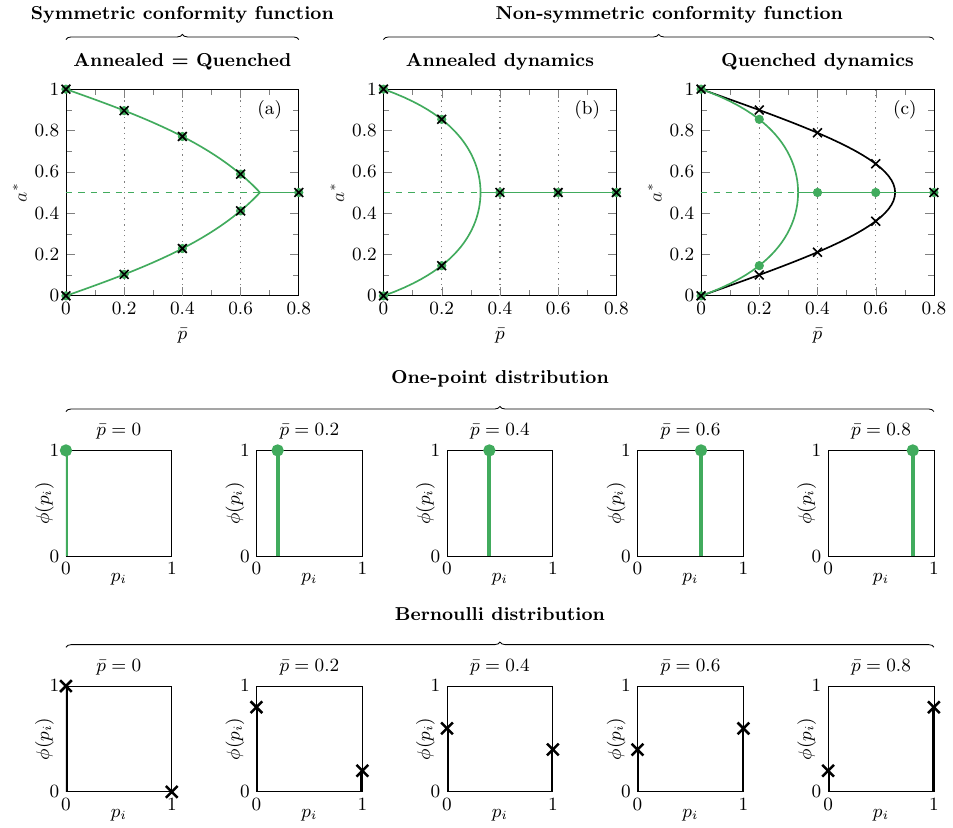}
	\caption{\label{fig:fixed-points-dist} \AJ{Illustration of how the change of the learning strategy distribution, $\phi(p_i)$, impacts the fixed-point diagram for the model with different types of conformity function and dynamics. 
    We present the case where $q=3$ and $k=0$. 
    The fixed points are plotted as a function of the mean of either one-point distribution (green lines and dots) or Bernoulli distribution (black lines and crosses), $\bar{p}$. Lower part of the figure illustrates the chosen distributions for a few values of $\bar{p}$.
 }
	}
\end{figure*}

\subsection{Specific case}
\label{sec:case}
\AJ{
To illustrate our findings, we choose specific forms of the learning strategy distribution and the learning functions, which are commonly discussed in the literature \cite{Yan:etal:21,Jed:Szn:19}.
However, it is important to note that any other functions can also be used \LH{as our general result is independent of the particular form of these functions}.
Our obective here is to visualize the general results applicable to a broad range of models with various functions, rather than analyzing a specific model in detail:
}
\begin{itemize}
    \item \AJ{\textit{Learning strategy distribution:}} Let us consider a simple case where $p_i$ is Bernoulli distributed, so $\forall i=1,...,N $, $\AJ{\phi(p_i=1)}=p$ and $\AJ{\phi(p_i=0)}=1-p$, and $p$ is the mean of the distribution.
    Note that agents with $p_i=1$ certainly behave as individual learners, whereas those with $p_i=0$ certainly behave as social learners.
    \item \AJ{\textit{Individual learning function:}} Following the Rasch model \cite{Bon:Yan:Hee:20}, we use the logit function to define \AJ{$I(a)$}.
    Thus, the natural logarithm of the odds of choosing $A$ is proportional to the fraction of individuals favoring this option over the fraction $a_m$, which gives the equal probability to both options: 
    \begin{equation}
    \label{eq:logodds}
        \ln\left[\frac{\AJ{I(a)}}{1-\AJ{I(a)}}\right]=k(a-a_m).
    \end{equation}
    The parameter  $k$  accounts for the tendency and the strength of  the likelihood of choosing option $A$, whereas  $a_m$ is the midpoint of \AJ{$I(a)$}, i.e., $\AJ{I(a_m)}=1/2$, which we set here as $a_m=1/2$.
    The left hand side of Eq.~(\ref{eq:logodds}) can be interpreted as a trade-off between an individual's attitudes towards option $A$ and its adoption difficulties \cite{Byr:etal:16,Kai:Byr:Har:10}. 
    In our model, this trade-off depends explicitly on the number of adopters. 
    From Eq.~(\ref{eq:logodds}), we get the following form of the individual leaning probability:
    \begin{equation}
        \AJ{I(a)}=\frac{1}{1+e^{-k(a-a_m)}}.
    \label{eq:individual_learning}
    \end{equation}
    \item \AJ{\textit{Social learning function:}}
        For the symmetric case, and for the sake of comparison, we use the same form as in Ref.~\cite{Yan:etal:21}:
    \begin{equation}
    \label{eq:symm-cf}
        \AJ{S(x)}=
        \begin{cases}
        \frac{1}{2}(2x)^q & \text{if }0 \leq x < 0.5,\\
        1-\frac{1}{2}\left[2(1-x)\right]^q & \text{if }0.5 \leq x \leq 1.
         \end{cases}
    \end{equation}
    For the  non-symmetric conformity functions, we assume a simple mathematical form inspired by the  \textit{non-linear q-voter model} \cite{Cas:Mun:Pas:09,Jed:Szn:19,Jed:Szn:20}:
    \begin{equation}
    \label{eq:nonsymm-cf}
        \AJ{S(x)}=x^q.
    \end{equation}
    Both these functions are parameterized by $q$, which reflects their degree of non-linearity, where $q=1$ corresponds to a linear response.
\end{itemize}

\subsubsection{Annealed dynamics}
Each agent is assigned a particular learning strategy at each time step: individual learning ($p_i=1$) and social learning $(p_i=0)$ with probability $p$ and $1-p$, respectively.
To describe such a system, we  simply use equations from Section~\ref{sec:gen-ann} with $\bar{p}=p$ since $p$ is the mean of the Bernoulli distribution in this case.
Notice that for $k=0$ and a non-symmetric conformity function given by Eq.~(\ref{eq:nonsymm-cf}), we get a particular case of the non-linear noisy voter model \cite{Per:etal:18} or the $q$-voter model with independence \cite{Nyc:Szn:Cis:12}.

\subsubsection{Quenched dynamics} 
Each agent is assigned a particular learning strategy from the Bernoulli distribution only once, at the start of the dynamics.
This means that eventually, we have two groups of agents. 
One group consists of individual learners ($p_i=1$), and the other group consists of social learners ($p_i=0$).
In such a system, individual learners represent a fraction $p$ of the total population, wheres social learners represent the remaining fraction $1-p$.
Notice that for  $k=0$, we obtain a particular case presented in Ref.~\cite{Dim:Gar:12} or \cite{Jed:Szn:17} depending on the considered type of the conformity function, symmetric for the former and non-symmetric for the latter.

In this case, Eq.~(\ref{eq:apopulation_que}) becomes
\begin{equation}
    a=(1-p)a_0+pa_1,
\end{equation}
where $a_0$ and $a_1$ are the factions of individuals who favor $A$ among social learners  ($p_i=0$) and individual learners ($p_i=1$), respectively.
The rate equations resulting from Eqs.~(\ref{eq:rate-que}) and (\ref{eq:transition-rates-que}) are the following:
\begin{equation}
\begin{split}
    \frac{da_0}{dt}&=\AJ{S(a)}(1-a_0)-\AJ{S(1-a)}a_0,\\
    \frac{da_1}{dt}&=\AJ{I(a)}-a_1.
\end{split}
\end{equation}
Thus, the fixed points satisfy:
\begin{equation}
\label{eq:fixed-que-ber}
    p=\frac{a^*\left[\AJ{S(a^*)+S(1-a^*)}\right]-\AJ{S(a^*)}}{\AJ{I(a^*)}\left[\AJ{S(a^*)+S(1-a^*)}\right]-\AJ{S(a^*)}}.
\end{equation}

\subsubsection{Result illustration}
Figure~\ref{fig:fixed-points-k} illustrates how the final state of the model is influenced by different dynamics when using symmetric and non-symmetric conformity functions.
In this example, $q=3$ and $k=-15$. Negative $k$ values correspond to the situation where the probability of adoption through individual learning diminishes with the fraction of adopters and is therefore in competition with the conformity function. 
 For symmetric conformity functions, both quenched and annealed curves overlap across  the entire range of $\bar{p}$, see Fig.~\ref{fig:fixed-points-k:a}.
 However, when non-symmetric conformity functions are considered, the fixed points may differ for certain values of $\bar{p}$, see Fig.~\ref{fig:fixed-points-k:b}.
These differences can be substantial, leading to continuous phase transitions for the quenched dynamics and discontinuous for the annealed one, as shown in this example.
The takeaway here is that the type of the dynamics does not impact the final state of models with a symmetric conformity function.
In contrast, the dynamics type becomes crucial in determining the final state of models with a non-symmetric conformity function.

Figure~\ref{fig:fixed-points-dist} demonstrates how the learning strategy distribution, $\phi(p_i)$, impacts the final state of the model with different types of conformity functions and dynamics.
In this example, $q=3$ and $k=0$.
We compare two simple distributions parameterized by a single variable $p\in[0,1]$: a one-point distribution where $\phi(p_i=p)=1$ and a Bernoulli distribution given by $\phi(p_i=1)=p$ and $\phi(p_i=0)=1-p$.
Note that both distributions share the same mean, $\bar{p}=p$.
The lower part of Fig.~\ref{fig:fixed-points-dist} illustrates these distributions for a few values of $\bar{p}$.
As seen in Fig.~\ref{fig:fixed-points-dist:a}, if a model uses a symmetric conformity function, annealed and quenched dynamics lead to the same fixed points for a given value of $\bar{p}$ across different distributions, $\phi(p_i)$. 
Similarly, for models with a non-symmetric conformity function and annealed dynamics, the shape of $\phi(p_i)$ does not impact the final state, see Fig.~\ref{fig:fixed-points-dist:b}.
However, if a model uses a non-symmetric conformity function and quenched dynamics, the mean of the distribution alone is insufficient to determine the fixed points since different distributions sharing the same mean may lead to different results, see Fig.~\ref{fig:fixed-points-dist:c}.

\section{Discussion}
Numerous theoretical and empirical studies aim to understand  and classify diverse social learning strategies \cite{Ren:etal:11,Mce:etal:08, Boy:Ric:88, Mor:Lal:12, Cla:etal:14,Apl:etal:15, Cla:Bow:Whi:12,Cla:Whi:12, Eff:etal:08, Der:God:Ray:13,Mor:etal:12,Ken:etal:18,Lat:81,Lum:Wil:80}.
In simple models of frequency-dependent bias \cite{Boy:Ric:88}, where individuals are assumed to be disproportionately more likely to adopt the most common option, this probability of adoption is represented by a conformity function which is an increasing function, symmetric around its midpoint $\AJ{S(0.5)}=0.5$.
Some experimental studies measuring the relationship between the option frequency and the probability of adoption tend to support this result~\cite{Cla:etal:14,Mce:etal:08,Eff:etal:08}.
However, when individuals have the information about the pay-offs of others, they may adaptively bias social learning, which leads to non-symmetric forms of conformity functions \cite{Ken:etal:18,Mce:etal:08,Mor:etal:12}.
Non-symmetric representations of social influence also appear in the social impact theory \cite{Lat:81}, models of cultural transmissions \cite{Boy:Ric:88, Lum:Wil:80}, and models of opinion dynamics \cite{Jed:Szn:19}.
Having recognized this distinction between symmetric and non-symmetric social learning functions, we raised a question about the implications of using either type for the decision-making process.

To this end, we generalised a simple dynamical system model in which individuals use either individual or social learning strategies to make their decisions \cite{Yan:etal:21}.
The generalized version allows us to model a broader range of systems, where the merit of an option can also change with the adoption rate \cite{Dim:Gar:12,Bal:Pei:18}.
Additionally, it accounts for an arbitrary distribution of personal inclinations towards strategies and distinguishes between annealed and quenched dynamics \cite{Jed:Szn:20,Jed:Szn:17}, which represent different timescales at which individuals change their inclinations.

Our analytical study of the generalized model indicates that when a symmetric conformity function is used, certain details of the model become irrelevant for the determination of the fixed points, which represent the final adoption rates in the system. 
In this case:
\begin{itemize}
    \item Both quenched and annealed dynamics converge to identical fixed points, so the timescale of the dynamics is not important.
    \item These fixed points depend only on the mean of the learning strategy distribution, so the distribution shape does not matter.
\end{itemize}
In contrast, the same details may play a crucial role in models featuring a non-symmetric conformity function. While for annealed dynamics, the fixed points still depend only on the mean of the learning strategy distribution,  the contrary is true for the quenched dynamics where the whole  distribution of learning strategies enters in the determination of the fixed points of the dynamics.

Interestingly,  in Ref.~\cite{Yan:etal:21}, where a particular case of this model with symmetric conformity functions and quenched dynamics is studied, the authors compare the results for three different distributions of learning strategies finding that the fixed-point diagram is very different in the case of a rightly skewed distribution of learning strategies compared to the cases of uniform or  truncated normal distributions.  
Our results clarify this point: the differences they found are the consequence that the mean of the  right-skewed distribution they chose is simply different than the mean of  the other two distributions.

The use of different learning strategies by individuals is influenced by contextual nuances, developmental experiences, and inter-individual variations \cite{Ren:etal:11,Ken:etal:18,Mce:etal:08}.
Some people exhibit greater inclinations to use social information than others \cite{Ken:etal:18,Eff:etal:08}. 
Factors like age, social rank, or popularity may play a role.
Individuals also adaptively change their inclinations based on varying levels of uncertainty \cite{Ken:etal:18, Mor:etal:12, Hop:Lal:13}.
Limited information or unreliable knowledge tends to increase the likelihood of social learning.
Mood and social context can also influence the strategy choice \cite{Ren:etal:11,Lal:04}.
All these studies highlight that people change their personal inclinations towards learning strategies.
If these changes occur frequently compared to how often people change their options, as in the annealed dynamics, the final adoption rate depends on the mean inclination in the system rather than the specific distribution of these preferences. 
However, if the individuals are more persistent with their inclinations, as in the quenched dynamics, and a system features a non-symmetric conformity function, the actual distribution of learning strategies becomes crucial for the final adoption rate.
In this case, more effort should be put into modelling various possible distributions of learning strategies and estimating them in empirical studies.

\section{Conclusions}
In this article, we study analytically and numerically the  outcomes of a decision making-process in a population of agents who may choose to learn individually or socially. We consider both symmetric and non-symmetric conformity functions along with annealed and quenched dynamics, which represents different timescales at which individuals change their inclinations towards learning strategies.
We show that the choice between a symmetric or non-symmetric conformity function overrules other model details in determining the final state of the system.

These findings might have practical implications not only for  theoretical modeling but also for experimental studies.
Commonly, experimental protocols measure the relationship between the probability of choosing a given option and the fraction of its adopters in the population and then  different functions are tried to fit  such data \cite{Mce:etal:08,Cla:Bow:Whi:12, Cla:etal:14,Mor:etal:12}.
However, there is not necessarily a one-to-one correspondence between the psychological rule employed by individuals and the population-level pattern that this rule produces \cite{Ken:etal:18}.
Our results suggest different experimental approaches that may provide information about some properties of  the actual cognitive rule employed by individuals.
One alternative method involves comparing adoption rates for two experimental setups that implement quenched and annealed dynamics.
If there is a statistically significant difference in adoption rates for a given mean inclination towards individual learning, it may indicate a non-symmetric character of a conformity function.
The other method involves controlling the shape of the distribution of preferences towards a learning strategy in a sample. In this scenario, one can compare adoption rates for different distributions of these preferences having the same mean in a setup representing only quenched dynamics.
Note that these methods could potentially offer insights into the symmetry of a social learning function by comparing just two points.

\begin{acknowledgments}
This project has received funding from the European Union's Horizon 2020 research and innovation program under the Maria Sk\l{}odowska-Curie  grant agreement number 945380.
\end{acknowledgments}

\appendix
\onecolumngrid
\section{Analytical calculations for the specific case}
In the main text, we show that if the conformity function is symmetric, the shape of the distribution of the learning strategies does not enter into the fixed-point equations, and only the mean of this distribution matters.
Moreover, the fixed-point equation is the same for the annealed and quenched dynamics.
To illustrate this general result, we use some particular functions that describe the individual and social learning as well as the distribution of learning strategies.
Herein, we present the calculations for these particular cases.
\subsection{Symmetric conformity function}
We have chosen 
\begin{equation}
    \AJ{S(x)}=
    \begin{cases}
    \frac{1}{2}(2x)^q & \text{if }0 \leq x < 0.5,\\
    1-\frac{1}{2}\left[2(1-x)\right]^q & \text{if }0.5 \leq x \leq 1,
     \end{cases}
     \label{SI_symmetric_cf}
\end{equation}
as our representative of symmetric conformity functions in order to compare with the particular case studied in Ref.~\cite{Yan:etal:21}.
In the case of symmetric conformity functions, annealed and quenched dynamics lead to the same final result.
This result depends only on the mean of the distribution of learning strategies \AJ{$\phi(x)$}, i.e., $\bar{p}=\int x\AJ{\phi(x)}dx$.
However, since the calculations that lead to this result are different for annealed and quenched dynamics, we cover them separately in the next subsections.
\subsubsection{Annealed dynamics}
The rate equation takes the following form:
\begin{equation}
\label{eq:rate-eq-ann}
    \frac{da}{dt}=
    \begin{dcases}
    \bar{p}\left[\AJ{I(a)}-a\right]+(1-\bar{p})\left[\frac{1}{2}(2a)^q-a\right]  & \text{if }0 \leq a < 0.5,\\
    \bar{p}\left[\AJ{I(a)}-a\right]+(1-\bar{p})\left\{1-a-\frac{1}{2}\left[2(1-a)\right]^q\right\}  & \text{if }0.5 \leq a \leq 1.
     \end{dcases}
\end{equation}
We have two groups of fixed points. 
The first group is given by $a^*=1/2$ and any value of $\bar{p}$, whereas the second group satisfies the following formula:
\begin{equation}
\label{eq:fixed-ann-sym}
    \bar{p}=
    \begin{dcases}
    \frac{2a^*-(2a^*)^q}{2\AJ{I(a^*)}-(2a^*)^q}  & \text{if }0 \leq a^* < 0.5,\\
    \frac{2(a^*-1)+\left[2(1-a^*)\right]^q}{2\left[\AJ{I(a^*)}-1\right]+\left[2(1-a^*)\right]^q}  & \text{if }0.5 \leq a^* \leq 1.
     \end{dcases}
\end{equation}
To check the stability of the derived fixed points, let us first denote the right hand side of Eq.~(\ref{eq:rate-eq-ann}) by $F(a)$
and let
\begin{equation}
    F'(a^*)=\left.\frac{dF(a)}{da}\right\vert_{a^*}.
\end{equation}
The fixed point is stable if $F'(a^*)<0$ and unstable if $F'(a^*)>0$ \cite{Str:94}.
In this case,
\begin{equation}
    F'(a^*)=
    \begin{dcases}
    \bar{p}\left[\AJ{I'(a^*)}-1\right]+(1-\bar{p})\left[q(2a^*)^{q-1}-1\right]  & \text{if }0 \leq a^* < 0.5,\\
    \bar{p}\left[\AJ{I'(a^*)}-1\right]+(1-\bar{p})\left\{q\left[2(1-a^*)\right]^{q-1}-1\right\}  & \text{if }0.5 \leq a^* \leq 1,
    \end{dcases}
\end{equation}
where
\begin{equation}
    \AJ{I'(a^*)}=\left.\frac{d\AJ{I(a)}}{da}\right\vert_{a^*}=\frac{ke^{-k(a^*-1/2)}}{\left[1+e^{-k(a^*-1/2)}\right]^2}.
\end{equation}
For $a^*=1/2$, we can determine the stability analytically.
We have
\begin{equation}
    \AJ{I'(1/2)}=\frac{k}{4},
\end{equation}
so
\begin{equation}
    F'(1/2)=
    \frac{k}{4}\bar{p}+q(1-\bar{p})-1.
\end{equation}
Consequently, the point at which the stability of $a^*=1/2$ changes is given by
\begin{equation}
\label{eq:stab-ann-sym}
    \bar{p}^*=\frac{4\left(q-1\right)}{4q-k}.
\end{equation}
If $q>1$ and $k<4$, the fixed point $a^*=1/2$ is unstable for $\bar{p}<\bar{p}^*$, and it is stable for $\bar{p}>\bar{p}^*$, whereas if $k>4$, $a^*=1/2$ is unstable for all $\bar{p}$.
If $0<q<1$ and $k<4$, $a^*=1/2$ is stable for all $\bar{p}$, whereas if $k>4$, $a^*=1/2$ is stable for $\bar{p}<\bar{p}^*$, and it is unstable for $\bar{p}>\bar{p}^*$.
The stability of the remaining fixed points, given by Eq.~(\ref{eq:fixed-ann-sym}), is determined numerically.

\subsubsection{Quenched dynamics}
In this case, the rate equations have the following forms
\begin{align}
\label{eq:rate1-eq-que}
    \frac{da_0}{dt}&=
    \begin{dcases}
    \frac{1}{2}(2a)^q-a_0  & \text{if }0 \leq a < 0.5,\\
    1-a_0-\frac{1}{2}\left[2(1-a)\right]^q  & \text{if }0.5 \leq a \leq 1,
     \end{dcases}\\
     \label{eq:rate2-eq-que}
     \frac{da_1}{dt}&=\AJ{I(a)}-a_1, 
\end{align}
where
\begin{equation}
    a=(1-p)a_0+pa_1.
\end{equation}
The first group of fixed points is given by $(a_0^*,a_1^*)=(1/2,1/2)$ and any value of $\bar{p}$, whereas the second group satisfies the following formulas:
\begin{align}
    a_0^*&=
    \begin{dcases}
    \frac{1}{2}(2a^*)^q  & \text{if }0 \leq a^* < 0.5,\\
    1-\frac{1}{2}\left[2(1-a^*)\right]^q  & \text{if }0.5 \leq a^* \leq 1,
     \end{dcases}\\
     a_1^*&=\AJ{I(a^*)},
\end{align}
where 
\begin{equation}
a^*=(1-\bar{p})a_0^*+\bar{p}a_1^*.
\end{equation}
Consequently, we have
\begin{equation}
\label{eq:fixed-que-sym}
    \bar{p}=
    \begin{dcases}
    \frac{2a^*-(2a^*)^q}{2\AJ{I(a^*)}-(2a^*)^q}  & \text{if }0 \leq a^* < 0.5,\\
    \frac{2(a^*-1)+\left[2(1-a^*)\right]^q}{2\left[\AJ{I(a^*)}-1\right]+\left[2(1-a^*)\right]^q}  & \text{if }0.5 \leq a^* \leq 1.
     \end{dcases}
\end{equation}
Note that the same result was obtained for the annealed dynamics, see Eq.~(\ref{eq:fixed-ann-sym}).

To check the stability of the derived fixed points, let us denote the right hand side of Eqs.~(\ref{eq:rate1-eq-que}) and (\ref{eq:rate2-eq-que}) by $F_0(a_0,a_1)$ and $F_1(a_0,a_1)$, respectively.
The stability is determined by the determinant and trace of the following Jacobian matrix:
\begin{equation}
\label{eq:jac-matrix}
    \mathbf{J}(a_0^*,a_1^*)=\begin{bmatrix}
\dfrac{\partial F_0}{\partial a_0} & \dfrac{\partial F_0}{\partial a_1}\\[1em]
\dfrac{\partial F_1}{\partial a_0} & \dfrac{\partial F_1}{\partial a_1}
\end{bmatrix}_{(a_0^*,a_1^*)},
\end{equation}
where
\begin{align}
    \frac{\partial F_0}{\partial a_0}&=
    \begin{dcases}
    q(1-\bar{p})(2a)^{q-1}-1 & \text{if }0 \leq a < 0.5,\\
    q(1-\bar{p})\left[2(1-a)\right]^{q-1}-1 & \text{if }0.5 \leq a \leq 1,
    \end{dcases}\\
    \frac{\partial F_0}{\partial a_1}&=\begin{dcases}
    q\bar{p}(2a)^{q-1} & \text{if }0 \leq a < 0.5,\\
    q\bar{p}\left[2(1-a)\right]^{q-1} & \text{if }0.5 \leq a \leq 1,
    \end{dcases}\\
    \frac{\partial F_1}{\partial a_0}&=(1-\bar{p})\AJ{I'(a)},\\
    \frac{\partial F_1}{\partial a_1}&=\bar{p}\AJ{I'(a)}-1.
\end{align}
The state is stable if $\det\left[\mathbf{J}(a_0^*,a_1^*)\right]>0$ and $\text{tr}\left[\mathbf{J}(a_0^*,a_1^*)\right]<0$ \cite{Str:94}.

For $(a_0^*,a_1^*)=(1/2,1/2)$, we can determine the stability analytically.
In this case, we have
\begin{align}
    \left.\frac{\partial F_0}{\partial a_0}\right\vert_{(1/2,1/2)}&=q(1-\bar{p})-1,\\
    \left.\frac{\partial F_0}{\partial a_1}\right\vert_{(1/2,1/2)}&=q\bar{p},\\
    \left.\frac{\partial F_1}{\partial a_0}\right\vert_{(1/2,1/2)}&=\frac{k}{4}(1-\bar{p}),\\
    \left.\frac{\partial F_1}{\partial a_1}\right\vert_{(1/2,1/2)}&=\frac{k}{4}\bar{p}-1.
\end{align}
Hence, the determinant and the trace are the following:
\begin{align}
    \det\left[\mathbf{J}(1/2,1/2)\right]&=1-\frac{k}{4}\bar{p}-q(1-\bar{p}),\\
    \text{tr}\left[\mathbf{J}(1/2,1/2)\right]&=\frac{k}{4}\bar{p}+q(1-\bar{p})-2=-\left(\det\left[\mathbf{J}(1/2,1/2)\right]+1\right).
\end{align}
As a result, the point at which the stability of $(a_0^*,a_1^*)=(1/2,1/2)$ changes is given by
\begin{equation}
    \bar{p}^*=\frac{4\left(q-1\right)}{4q-k},
\end{equation}
which is the same formula as obtained for the annealed dynamics, see Eq.~(\ref{eq:stab-ann-sym}), and we have the same stability conditions.
The stability of the remaining fixed points, given by Eq.~(\ref{eq:fixed-que-sym}), is determined numerically.

\subsubsection{Results}
\begin{figure}[t!]
\subfloat{\label{fig:sym:a}}
\subfloat{\label{fig:sym:b}}
\subfloat{\label{fig:sym:c}}
\subfloat{\label{fig:sym:d}}
\subfloat{\label{fig:sym:e}}
\subfloat{\label{fig:sym:f}}
\centering
	\includegraphics[width=\linewidth]{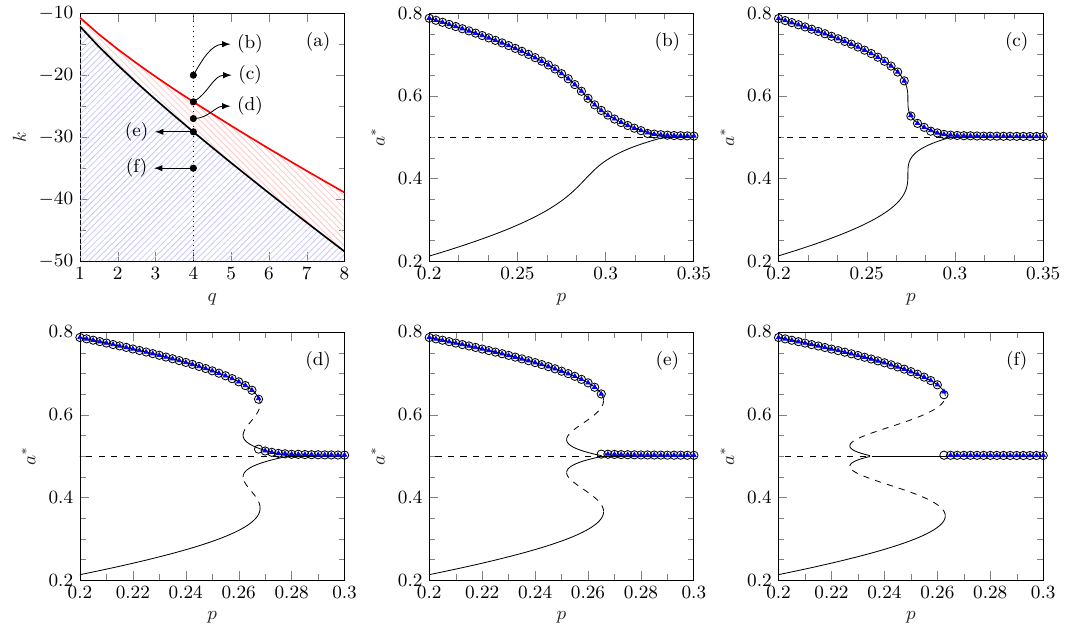}
	\caption{\label{fig:sym} Behavior of the model with symmetric conformity function, where the annealed and quenched dynamics produce the same diagrams. (a) Phase diagram. The blue and the white regions correspond to the zones of the parameter space where transitions between an ordered and a disordered phase are discontinuous or continuous, respectively. The intermediate red region also presents discontinuous transitions but  between two ordered phases with different fraction of adopters. The letters indicate the parameter regions of the following fixed-point diagrams (b)-(f), which present stable (solid lines) and unstable (dashed lines) fixed points for the model with $q=4$ and (b) $k=-20$, (c) $k=\widetilde{k}(q=4)\approx-24.3$, (d) $k=-27$, (e) $k=\bar{k}(q=4)\approx-29.1$, (f) $k=-35$. Symbols represent the results from the simulations of the model with $N=10^5$ agents under $\circ$ annealed and $\blacktriangle$ quenched dynamics.}
\end{figure}

Figure~\ref{fig:sym} illustrates the behavior of the model with the symmetric conformity function. In this case, the annealed and quenched dynamics produce the same fixed-point diagrams.
In the parameter space presented in Fig.~\ref{fig:sym:a}, we can identify three areas separated by two curves: $\widetilde{k}(q)$, the red one, and $\bar{k}(q)$, the black one.
These curves are determined numerically.
For $k>\widetilde{k}(q)$, the system exhibits continuous transitions between a phase where one option dominates over the other (i.e., ordered phase for $\bar{p}<\bar{p}^*$) to a phase without the majoritarian option (i.e., disordered phase for $\bar{p}>\bar{p}^*$), see Fig.~\ref{fig:sym:b}.
For $\bar{k}(q)<k<\widetilde{k}(q)$, additional discontinuous transitions between phases with the majoritarian options appear, see Fig.~\ref{fig:sym:d}.
Finally, for $k<\bar{k}(q)$, discontinuous transitions between phases with and without the majoritarian options are possible, see Fig.~\ref{fig:sym:f}.
\clearpage

\subsection{Non-symmetric conformity function}
We have chosen  
\begin{equation}
\label{eq:sym-cfun}
    \AJ{S(x)}=x^q
\end{equation}
as our representative of non-symmetric conformity functions as this form is commonly used in models of opinion dynamics \cite{Jed:Szn:19} that originate from the nonlinear $q$-voter model \cite{Cas:Mun:Pas:09}.
In the case of non-symmetric conformity functions, annealed and quenched dynamics lead to different results.
We cover them separately in the next subsections.

\subsubsection{Annealed dynamics}
The transition rates take the forms:
\begin{align}
\label{eq:transition-rates-nonsym-ann}
    P_{B\rightarrow A}&=\bar{p}\AJ{I(a)}+(1-\bar{p})a^q,\\
    P_{A\rightarrow B}&= \bar{p}\left[1-\AJ{I(a)}\right]+(1-\bar{p})(1-a)^q,
\end{align}
which results in the following rate equation
\begin{equation}
\label{eq:rate-eq-ann-nonsym}
    \frac{da}{dt}=\bar{p}\left[\AJ{I(a)}-a\right]+(1-\bar{p})\left[(1-a)a^q-a(1-a)^q\right].
\end{equation}
The first group of fixed points is given by $a^*=1/2$ and any value of $\bar{p}$, and the second group satisfies the following formula:
\begin{equation}
\label{eq:fixed-ann-voter}
    \bar{p}=\frac{a^*(1-a^*)^q-(1-a^*)(a^*)^q}{a^*\left[(a^*)^q+(1-a^*)^q-1\right]+\AJ{I(a^*)}-(a^*)^q}.
\end{equation}

To check the stability of the derived fixed points, let us denote the right hand side of Eq.~(\ref{eq:rate-eq-ann-nonsym}) by $F(a)$.
The stability is determined by the sign of
\begin{equation}
    F'(a^*)=\bar{p}\left[\AJ{I'(a^*)}-1\right]+(1-\bar{p})\left[q(1-a^*)(a^*)^{q-1}+qa^*(1-a^*)^{q-1}-(a^*)^q-(1-a^*)^q\right],
\end{equation}
where
\begin{equation}
    \AJ{I'(a^*)}=\frac{ke^{-k(a^*-a_0)}}{\left[1+e^{-k(a^*-a_0)}\right]^2}.
\end{equation}
For $a^*=1/2$, we can determine the stability analytically.
In this case, we have
\begin{equation}
    \AJ{I'(1/2)}=\frac{k}{4},
\end{equation}
and
\begin{equation}
    F'(1/2)=\bar{p}\left[\frac{k}{4}-1\right]+(1-\bar{p})\frac{q-1}{2^{q-1}}.
\end{equation}
Consequently, the point at which the stability of $a^*=1/2$ changes is given by
\begin{equation}
    \bar{p}^*=\frac{q-1}{q-1+2^{q-1}\left(1-\frac{k}{4}\right)}.
\end{equation}
If $q>1$ and $k<4$, the fixed point $a^*=1/2$ is unstable for $\bar{p}<\bar{p}^*$, and it is stable for $\bar{p}>\bar{p}^*$, whereas if $k>4$, $a^*=1/2$ is unstable for all $\bar{p}$.
If $0<q<1$ and $k<4$, $a^*=1/2$ is stable for all $\bar{p}$, whereas if $k>4$, $a^*=1/2$ is stable for $\bar{p}<\bar{p}^*$, and it is unstable for $\bar{p}>\bar{p}^*$.
The stability of the remaining fixed points, given by Eq.~(\ref{eq:fixed-ann-voter}), is determined numerically.

At the fixed point $(a^*,\bar{p})=(1/2,\bar{p}^*)$, a pitchfork bifurcation takes place. 
This bifurcation changes its type from subcritical to supercritical in the parameter space $(k,q)$ along the curve $k^*(q)$ defined by the equation:
\begin{equation}
\label{eq:ks-ann}
    (k^*)^3+8(k^*-4)(q-5)q=0.
\end{equation}
The bifurcation is subcritical for $k<k^*(q)$, while it becomes supercritical for $k>k^*(q)$.

Note that this model for $k=0$ corresponds to the $q$-voter model with independence \cite{Nyc:Szn:Cis:12, Nyc:Szn:13, Jed:Szn:17} or the non-linear noisy voter model \cite{Per:etal:18}.

\subsubsection{Quenched dynamics}
In this case, the rate equations have the following forms:
\begin{align}
    \label{eq:rate1-eq-que-nonsym}
    \frac{da_0}{dt}&=(1-a_0)a^q-a_0(1-a)^q,\\
    \label{eq:rate2-eq-que-nonsym}
    \frac{da_1}{dt}&=\AJ{I(a)}-a_1,
\end{align}
where
\begin{equation}
    a=(1-p)a_0+pa_1.
\end{equation}
The first group of fixed points is given by $(a_0^*,a_1^*)=(1/2,1/2)$ and any value of $\bar{p}$, whereas the second group satisfies the following formulas:
\begin{align}
    a_0^*&=\frac{(a^*)^q}{(a^*)^q+(1-a^*)^q},\\
    a_1^*&=\AJ{I(a^*)},
\end{align}
where 
\begin{equation}
a^*=(1-\bar{p})a_0^*+\bar{p}a_1^*.
\end{equation}
As a result, we have
\begin{equation}
\label{eq:fixed-que-ber-voter}
    \bar{p}=\frac{a^*(1-a^*)^q-(1-a^*)(a^*)^q}{\AJ{I(a^*)}\left[(a^*)^q+(1-a^*)^q\right]-(a^*)^q}.
\end{equation}

To check the stability of the derived fixed points, let us denote the right hand side of Eqs.~(\ref{eq:rate1-eq-que-nonsym}) and (\ref{eq:rate2-eq-que-nonsym}) by $F_0(a_0,a_1)$ and $F_1(a_0,a_1)$, respectively.
The stability is determined by the use of the Jacobian matrix given by Eq.~(\ref{eq:jac-matrix}) where
\begin{align}
    \frac{\partial F_0}{\partial a_0}&=q(1-\bar{p})\left[a_0(1-a)^{q-1}+(1-a_0)a^{q-1}\right]-a^q-(1-a)^q,\\
    \frac{\partial F_0}{\partial a_1}&=q\bar{p}\left[a_0(1-a)^{q-1}+(1-a_0)a^{q-1}\right],\\
    \frac{\partial F_1}{\partial a_0}&=(1-\bar{p})\AJ{I'(a)},\\
    \frac{\partial F_1}{\partial a_1}&=\bar{p}\AJ{I'(a)}-1.
\end{align}
For $(a_0^*,a_1^*)=(1/2,1/2)$, we can determine the stability analytically.
In this case, we have
\begin{align}
    \left.\frac{\partial F_0}{\partial a_0}\right\vert_{(1/2,1/2)}&=\frac{1}{2^{q-1}}\left[q(1-\bar{p})-1\right],\\
    \left.\frac{\partial F_0}{\partial a_1}\right\vert_{(1/2,1/2)}&=\frac{1}{2^{q-1}}q\bar{p},\\
    \left.\frac{\partial F_1}{\partial a_0}\right\vert_{(1/2,1/2)}&=\frac{k}{4}(1-\bar{p}),\\
    \left.\frac{\partial F_1}{\partial a_1}\right\vert_{(1/2,1/2)}&=\frac{k}{4}\bar{p}-1.
\end{align}
Thus, the determinant and the trace are the following:
\begin{align}
    \det\left[\mathbf{J}(1/2,1/2)\right]&=\frac{1}{2^{q-1}}\left[1-\frac{k}{4}\bar{p}-q(1-\bar{p})\right],\\
    \text{tr}\left[\mathbf{J}(1/2,1/2)\right]&=\frac{1}{2^{q-1}}\left[q(1-\bar{p})-1\right]+\frac{k}{4}\bar{p}-1.
\end{align}
As a result, the point at which the stability of $(a_0^*,a_1^*)=(1/2,1/2)$ changes is
\begin{equation}
    \bar{p}^*=\frac{4\left(q-1\right)}{4q-k}.
\end{equation}
If $q>1$ and $k<4$, the fixed point $(a_0^*,a_1^*)=(1/2,1/2)$ is unstable for $\bar{p}<\bar{p}^*$, and it is stable for $\bar{p}>\bar{p}^*$, whereas if $k>4$, $(a_0^*,a_1^*)=(1/2,1/2)$ is unstable for all $\bar{p}$.
If $0<q<1$ and $k<4$, $(a_0^*,a_1^*)=(1/2,1/2)$ is stable for all $\bar{p}$, whereas if $k>4$, $(a_0^*,a_1^*)=(1/2,1/2)$ is stable for $\bar{p}<\bar{p}^*$, and it is unstable for $\bar{p}>\bar{p}^*$.
The stability of the remaining fixed points, given by Eq.~(\ref{eq:fixed-que-ber-voter}), is determined numerically.

At the fixed point $(a^*,\bar{p})=(1/2,\bar{p}^*)$, a pitchfork bifurcation takes place. 
This bifurcation changes its type from subcritical to supercritical in the parameter space $(k,q)$ along the curve $k^*(q)$ defined by the equation:
\begin{equation}
\label{eq:ks-que}
    (k^*)^3+16(4-k^*)(q+1)q=0
\end{equation}
The bifurcation is subcritical for $k<k^*(q)$, while it becomes supercritical for $k>k^*(q)$.

Note that this model for $k=0$ corresponds to the $q$-voter model with independence under the quenched approach from Ref.~\cite{Jed:Szn:17}.

\subsubsection{Results}
\begin{figure}[t!]
\subfloat{\label{fig:nonsym:a}}
\subfloat{\label{fig:nonsym:b}}
\subfloat{\label{fig:nonsym:c}}
\subfloat{\label{fig:nonsym:d}}
\subfloat{\label{fig:nonsym:e}}
\subfloat{\label{fig:nonsym:f}}
\subfloat{\label{fig:nonsym:g}}
\subfloat{\label{fig:nonsym:h}}
\centering
	\includegraphics[width=\linewidth]{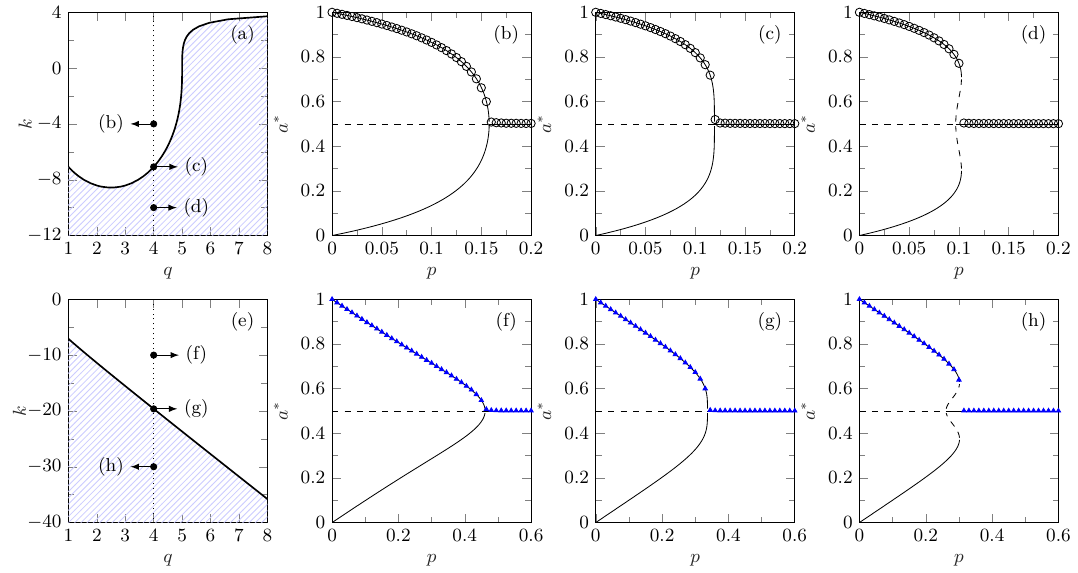}
	\caption{\label{fig:nonsym} Behavior of the model with non-symmetric conformity function for (a)-(d) annealed and (e)-(h) quenched dynamics. 
  (a) and (e) Phase diagrams. The blue and the white regions correspond to the zones of the parameter space where transitions between an ordered and a disordered phase are discontinuous or continuous, respectively.
  The letters indicate the parameter regions of the following fixed-point diagrams (b)-(d) and (f)-(h), which present stable (solid lines) and unstable (dashed lines) fixed points for the model with $q=4$ and (b) $k=-4$, (c) $k=k^*(q=4)\approx-7.1$, (d) $k=-10$, (f) $k=-10$, (g) $k=k^*(q=4)\approx-19.6$, (h) $k=-30$. Symbols represent the results from the simulations of the model with $N=10^5$ agents under $\circ$ annealed and $\blacktriangle$ quenched dynamics.}
\end{figure}

Figure~\ref{fig:nonsym} illustrates the behavior of the model with the non-symmetric conformity function for (a)-(d) annealed and (e)-(h) quenched dynamics.
In the parameter space presented in Fig.~\ref{fig:nonsym:a} and \ref{fig:nonsym:e}, we can identify two areas separated by the black curve, $k^*(q)$, given by Eqs.~(\ref{eq:ks-ann}) and (\ref{eq:ks-que}).
For $k>k^*(q)$, the system exhibits continuous transitions between a phase where one option dominates over the other (i.e., ordered phase for $\bar{p}<\bar{p}^*$) to a phase without the majoritarian option (i.e., disordered phase for $\bar{p}>\bar{p}^*$), see Figs.~\ref{fig:nonsym:b} and \ref{fig:nonsym:f}.
At $k^*(q)$, the system still exhibits continuous phase transitions, see Figs.~\ref{fig:nonsym:c} and \ref{fig:nonsym:g}. However, crossing this curve results in a  change of the phase transition type. 
Consequently, for $k<k^*(q)$, the transitions between ordered and disordered phases are discontinuous, see Figs.~\ref{fig:nonsym:d} and \ref{fig:nonsym:h}.
\clearpage
\section{Simulations}
\label{sec:simulations}
\subsection{Simulation details}
In addition to the analytical results discussed in the main text, which correspond to the thermodynamic limit, we simulate the dynamical equations for a large but finite population of $N=10^5$ agents.
We consider one time step of this dynamics when $N$ agents have been updated, or in other words when, on average, all the agents have been updated once, in analogy with the notion of \textit{Monte Carlo step per site} (MCS/s).
In the simulations, we trace the fraction of adopters of the most common option, i.e.,
\begin{equation}
    \alpha=\max\{a,b\},
\end{equation}
where $a$ and $b=1-a$ are the fraction of adopters of the options $A$ and $B$, respectively.
In the figures, we show the mean value of $\alpha$, $\left[\langle\alpha\rangle_t\right]_s$. The angle brackets represent the average over time. 
We discarded the first 900~MCS to let the system reach the stationary state and perform the time average over next 100~MCS. 
The square brackets represent the sample average that was performed over 20 independent simulations at most (in a metastable region, the average is perform over those of the simulations that ended up in the same phase).
In all the simulations, all the agents are initialized with option $A$.
Standard errors are of the mark size order.

When simulating the quenched dynamics, instead of randomly assigning the learning strategies, which would lead to some fluctuations in $\bar{p}$ between simulations, we assign them deterministically.
We choose the first $pN$ agents to be individual learners and the rest of them to be social learners. In such a way, we have exactly the same value of $\bar{p}$ in all the simulations that we average over.
The problem with the fluctuations can be also overcome by keeping the random assignment but increasing the number of agents in the system as the fluctuations in $\bar{p}$ diminishes with the system size at a rate of $1/\sqrt{N}$.

\subsection{Source code}
The model is implemented in C++ using object-oriented programming. 
Python and Matlab are used for data analysis and numerical calculations.
The code files can be found in the following GitHub repositories:
\begin{itemize}
    \item \href{https://github.com/arkadiusz-jedrzejewski/norm-formation-abm}{https://github.com/arkadiusz-jedrzejewski/norm-formation-abm},
    \item \href{https://github.com/arkadiusz-jedrzejewski/norm-formation-abm-py}{https://github.com/arkadiusz-jedrzejewski/norm-formation-abm-py},
    \item \href{https://github.com/arkadiusz-jedrzejewski/norm-formation-m}{https://github.com/arkadiusz-jedrzejewski/norm-formation-m}.
\end{itemize}
\twocolumngrid
\bibliography{main}

\end{document}